\documentclass[twocolumn,superscriptaddress,floatfix,aps,prl]{revtex4}
\usepackage{graphicx,subfigure,epsfig}
\usepackage{dcolumn}
\usepackage{amssymb}
\usepackage{times}
\usepackage{amsmath}
\usepackage{amsfonts}
\usepackage{mathrsfs}
\usepackage{setspace}
\usepackage{latexsym}
\usepackage{bbm}
\usepackage{float}
\usepackage{flafter}
\usepackage{bm}
\usepackage{epstopdf}
\usepackage{hyperref}
\usepackage{color}
\usepackage{multirow}

\begin{document}
\title{Magnetic Field Induced Spin Liquids in S=1 Kitaev Honeycomb Model}

\author{Zheng Zhu}
\email[\href{mailto:zhuzhengphysics@gmail.com}{zhuzhengphysics@gmail.com}]{}
\affiliation{Department of Physics, Harvard University, Cambridge, MA, 02138, USA}
\author{Zheng-Yu Weng}
\affiliation{Institute for Advanced Study, Tsinghua University, Beijing, 100084, China}
\author{D. N. Sheng}
\affiliation{Department of Physics and Astronomy, California State University, Northridge, CA, 91330, USA}

\begin{abstract}
We investigate the ground state properties of the spin S=1 Kitaev honeycomb model under a magnetic field based on the density matrix renormalization group (DMRG) calculation. With the time reversal symmetry breaking due to the magnetic field, a gapped Kitaev spin liquid is identified for both ferromagnetic (FM) and antiferromagnetic (AFM) Kitaev couplings. The topological nature of such Kitaev spin liquid is  manifested by the nearly quantized Wilson loop, degeneracy in the entanglement spectra  and existence of edge modes. While the FM Kitaev spin liquid is destroyed by a weaker magnetic field $H_*^\text{FM}$, the AFM one demonstrates a robustness up to an order of magnitude larger critical field $H_*^\text{AFM}$. Moreover, an intermediate nonmagnetic phase appears only for the AFM case at larger fields, $H_*^\text{AFM} < H < H_{**}^\text{AFM}$, before the transition to a high-field polarized paramagnet.
The stability of the Kitaev spin liquid against the Heisenberg interactions is also examined. Our findings may further inspire the investigation of recently proposed S=1 Kitaev materials.
\end{abstract}
\maketitle

\begin{figure}[t]
\begin{center}
\includegraphics[width=0.5\textwidth]{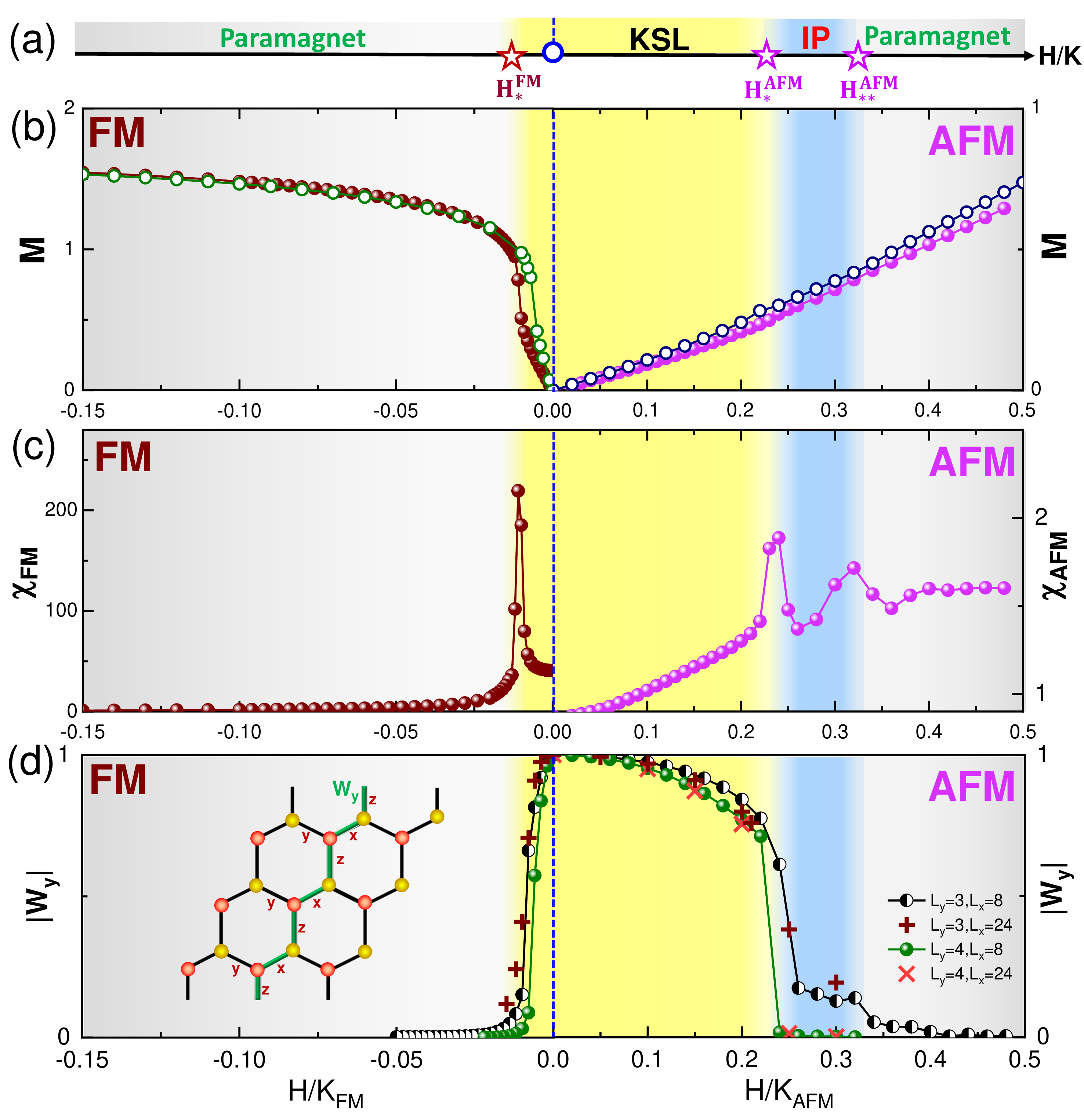}
\end{center}
\par
\renewcommand{\figurename}{Fig.}
\caption{(Color online) Phase diagram of S=1 Kitaev honeycomb model subject to a [111] magnetic field $H$.  Panel (a) shows the distinct phases in FM (left side) and AFM (right side) Kitaev model as a function of magnetic field strength, including the gapped Kitaev spin liquid (KSL) phase, the intermediate phase (IP), and the partial polarized paramagnet. The zero field limit (blue circle) corresponds to the pure Kitaev model. The phase boundaries are probed by the magnetization $M$ (b) and the magnetic susceptibility $\chi$ (c) for systems on torus with $4\times3$ unit cells (sphere), and on cylinder with $8\times 4$ (circle) .  The peaks in $\chi$ suggest the phase transitions. Note the stark difference between the AFM and FM models,  $H_*^\text{AFM}$ $\gg$ $H_*^\text{FM}$.  Panel (c) shows the Wilson loop $W_y$ (illustrated by the green loop in the inset) as a function of $H$, the nearly quantized $W_y$ implies the approximate $\mathbb{Z}_2$ gauge structure in KSL phase.}
\label{Fig:PhaseDiagram}
\end{figure}
\begin{figure*}[t]
\begin{center}
\includegraphics[width=1\textwidth]{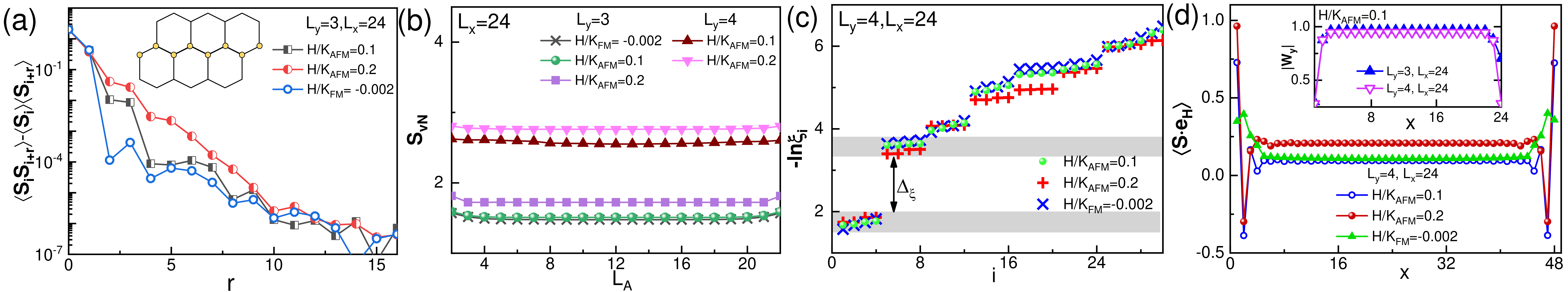}
\end{center}
\par
\renewcommand{\figurename}{Fig.}
\caption{(Color online) Magnetic field induced Kitaev spin liquid in the AFM/FM Kitaev models under the [111] magnetic field.  Panel (a) shows the decay of spin correlations along the zigzag chain (the yellow dots in the inset) in semi-logarithmic scale.
The gapped nature of KSL can also be confirmed from the Von Neumann entanglement entropy $S_\text{VN}$ in panel (b), which is independent of the cut position $L_A$ (the cuts are parallel to $\mathbf{e_y}$). The corresponding entanglement spectra at $L_A=L_x/2$ is shown in panel (c), which show four quasi-degenerate states separated from higher spectra by a finite entanglement gap $\Delta_\xi$. In panel (d),  the magnetization distribution $\langle\mathbf{S}\cdot\mathbf{e_{\text{H}}}\rangle$ [$\mathbf{e_{\text{H}}}=(1,1,1)$] along the zigzag chain explicitly exhibits the edge modes. The inset of (d) shows the uniform distribution of Wilson loop $W_y$ along the cylinder.
}
\label{Fig:GappedSL}
\end{figure*}

 \emph{Introduction.}---The search for exotic states of matter such as highly entangled quantum spin liquids (QSLs) has attracted considerable attention in modern condensed matter field~\cite{Lee2008,Balents,Zhou2017,Knolle2019}.  Among various proposed QSLs, the Kitaev model~\cite{Kitaev2006} and its candidate materials~\cite{Jackeli2009, Chaloupka2010, Rau2016,Trebst2017,Hermanns2017} are of particular interest.

The Kitaev model on a  tri-coordinated lattice with S=1/2 local degree of freedom is exactly solvable by mapping it into a model of Majorana fermions coupled to $\mathbb{Z}_2$ gauge fields~\cite{Kitaev2006}. The ground state harbors both a gapless QSL for a parameter region around equal coupling strengths and a gapped QSL for more anisotropic couplings. Interestingly,  the former becomes a non-Abelian topological spin liquid upon time-reversal symmetry breaking~\cite{Kitaev2006}, for example by applying a magnetic field~\cite{Kitaev2006}, adding a three-spin chirality term~\cite{DHLee2007} or decorating the honeycomb lattice~\cite{Yao2007}.
The realization of Kitaev physics in real materials has been proposed in transition metal oxides with strong spin-orbit coupling (SOC) and $S=1/2$ local moments, in which the edge-sharing oxygen octahedra result in bond-dependent Ising interactions~\cite{Jackeli2009,Chaloupka2010}. By now, various materials including  A$_2$IrO$_3$ (A=Na, Li)~\cite{Singh,Choi2012}, $\alpha$-RuCl$_3$~\cite{Plumb2014} and H$_3$LiIr$_2$O$_6$\cite{Takagi2017} have been discovered to be proximate to Kitaev spin liquid (KSL) \cite{Jiang2011,Katukuri2014,Kimchi2014,Iregui2014,Gohlke2017,Banerjee2016,Winter2017,Janssen2017,Ponomaryov2017,Little2017,ZWang2017, Song2016,Kasahara2018}.

Remarkably, the recent observation of half-integer quantized thermal Hall conductivity in $\alpha$-RuCl$_3$ under magnetic field offers a smoking-gun evidence of the fractionalized topological state~\cite{Kasahara2018}, leading to substantial activities in examining the magnetic field induced QSLs in the Kitaev models and materials~\cite{ZZ2018,Gohlke2018,Nasu2018,Liang2018,Hickey2019,Trivedi,Jacob2019,Jiang2018,Jiang2019,Takikawa2019, Liu2018, Janssen2019,Ido2019,Jang2019,Li2019,Chen2019,Wen2019,Gao2019,Natori2019,Kaib2019,Lee2019a,Sugita2019}. Although numerous efforts have been devoted to S=1/2 case,
the magnetic field induced phases in the S=1 Kitaev model and the corresponding experiments are of equal interest and importance.
Different from S=1/2, the nature of the ground state for the pure S=1 Kitaev model is still controversial~\cite{Baskaran,Minakawa2019,ED,TPQ,DMRG,Lee2019}.
Recently, a mechanism for realizing the S=1 Kitaev interactions by considering both SOC and strong Hund's couplings in transition metal oxides A$_3$Ni$_2$XO$_6$ (A=Li, Na and X=Bi, Sb) has been proposed~\cite{Stavropoulos2019}, providing a platform to explore the S=1 Kitaev physics.  The honeycomb layer lies in a plane perpendicular to [111] direction and the dominant Kitaev coupling is antiferromagnetic~\cite{Stavropoulos2019}. It is then natural to ask if the perpendicular magnetic field can induce new QSLs in such real materials, which is distinct from the pure S=1 Kitaev model ~\cite{Baskaran,Minakawa2019,ED,TPQ,DMRG,Lee2019} due to the lack of time reversal symmetry~\cite{Kitaev2006}.  Here, both the possible field-induced phases and their instabilities are equally important to the ones in S=1/2 case~\cite{ZZ2018,Gohlke2018,Nasu2018,Liang2018,Hickey2019,Trivedi,Jacob2019,Jiang2018,Jiang2019,Takikawa2019, Liu2018, Janssen2019,Ido2019,Jang2019,Li2019,Chen2019,Wen2019,Gao2019,Natori2019,Kaib2019,Lee2019a,Sugita2019}.Therefore, establishing the quantum phase diagram is the very essence of understanding various important questions and  providing valuable information for the experiments. Motivated by the above, in this Rapid Communication we study the magnetic field induced phases in the S=1 Kitaev honeycomb model by DMRG.

\emph{Model, Method and Key Results.}---We consider the S=1 Kitaev honeycomb model with a perpendicular magnetic field $\mathbf{H}$ along [111] direction. The Hamiltonian is
\begin{equation}\label{Model}
H=\sum\limits_{\left\langle {i,j} \right\rangle } {{K_\gamma }S_i^\gamma S_j^\gamma }-\sum\limits_i {{\mathbf{H}}\cdot{{\mathbf{S}}_i}}+ J\sum\limits_{\left\langle {i,j} \right\rangle } \mathbf{S}_i\cdot\mathbf{S}_j.
\end{equation}
Here, $\gamma=x,y,z$ represent the three nearest-neighbor links $\langle {i,j}\rangle$ of the honeycomb lattice, $S^\gamma$ denotes effective spin S=1 degrees of freedom sitting on each site and interacting via Kitaev coupling $K_\gamma$.  The second term is the Zeeman term with a uniform field,  the third term represents the Heisenberg interactions with coupling $J$.  We consider isotropic coupling $K_\gamma\equiv K$ and set $|K|$ as the unit of energy,  $K=+1$ ($K=-1$) corresponds to AFM (FM) model. Below we will identify the magnetic field induced phases at $J=0$ and examine the stability of KSL against Heisenberg perturbations.

We use DMRG to study the Hamiltonian (\ref{Model}) on a honeycomb lattice spanned by length vectors $\mathbf{L_x}=L_x \mathbf{e_x}$ and $\mathbf{L_y}=L_y\mathbf{e_y}$, where $\mathbf{e_x}=(1,0)$ and $\mathbf{e_y}=(1/2, \sqrt{3}/2)$ are two primitive vectors.  As each unit cell contains two sites, and the total number of sites $N=L_x\times L_y\times 2$. In the present calculation, we choose both the cylinder and torus geometries and keep up to 1600$\sim$2800 states for good convergence (depending on the nature of states and system sizes). We keep  the truncation error to be smaller than $10^{-6}$ in most cases.

We shall identify a gapped KSL phase for both antiferromagnetic (AFM) and ferromagnetic (FM) Kiatev couplings in weak magnetic fields. Interestingly while the FM gapped KSL is fragile against the magnetic field, the AFM gapped KSL is extremely robust up to an order of magnitude larger critical field. We also find the topological nature of this gapped KSL, including the $\mathbb{Z}_2$ gauge structure, the edge modes and a robust multiple  degeneracy in the entanglement spectra on ladder systems. Moreover, we discover an intermediate phase (IP) only in the AFM model before a second transition to the high-field polarized paramagnet. The phase diagram is depicted in Fig.~\ref{Fig:PhaseDiagram} (a). Finally we show that field-induced KSLs eventually give way to magnetic orders under sufficient Heisenberg  perturbations.

\emph{Phase Diagram.}---We begin with establishing the phase diagram of the model Hamiltonian (\ref{Model}) [see Fig.~\ref{Fig:PhaseDiagram} (a)]. We obtain the ground state for a given magnetic field, and the phase boundaries are determined by consistent evidence from the measurement results of
magnetization $M$, magnetic susceptibility $\chi$ and the Wilson loop $W_y$.

Figures~\ref{Fig:PhaseDiagram} (b-c) show the magnetization $M$ and magnetic susceptibility $\chi=\partial M/\partial H$ for  systems with both toroidal and cylindrical boundary conditions. At the FM Kitaev coupling side [see the left-hand side in Fig.~\ref{Fig:PhaseDiagram}], we find a strong response of the system to magnetic field, the single kink in magnetization $M$  [see Fig.~\ref{Fig:PhaseDiagram} (b)] and single peak of $\chi$ [see Fig.~\ref{Fig:PhaseDiagram} (c)] demonstrate a single phase transition to the high-field partially-polarized paramagnet at $H_*^\text{FM}\sim 0.01$, which smoothly connects to the fully polarized phase in infinite  field limit. In contrast, in the case with AFM  Kitaev coupling [see the right-hand side in Fig.~\ref{Fig:PhaseDiagram}], the system exhibits a weak response to an external field with a critical field $H_*^\text{AFM}\sim 0.24$,  which is larger than  $H_*^\text{FM}$ by more than an order of magnitude [see Fig.~\ref{Fig:PhaseDiagram} (c)]. Moreover, the two-peak structure in the  magnetic susceptibility $\chi$ suggests that there is an intermediate phase at $H_*^\text{AFM} < H < H_{**}^\text{AFM}$ before a second transition to partial polarized paramagnet beyond $H_{**}^\text{AFM}$.

We also note that the ground state of S=1 Kitaev model at zero field can be characterized by a conserved $\mathbb{Z}_2$ quantum number determined by the Wilson loop operators $\hat{W}_l$, which is generalized from S=1/2 model and can be defined by $\hat{W}_l=\prod_{i =1}^{L_l} \text{exp}[i\pi S_i^{\alpha_i}]$ in every closed loop on the lattice, where $\alpha_i$ represents the normal direction of site $i$~\cite{Baskaran}.  For example, $\hat{W}_y$ denotes the Wilson loop operator with the loop winding once around the cylinder which only cover $\gamma=x,z$ links [see the inset in Fig.~\ref{Fig:PhaseDiagram} (d)]. It is straightforward to verify that these Wilson loop operators commute with each other and also commute with the Hamiltonian of the \emph{pure} Kitaev model. Since $\hat{W}_l$ squares to identity $\hat{W}_l^2=1$, the eigenvalues of $\hat{W}_l$ are $W_l=\pm 1$ at $H=0$, which is associated with flux excitations like in the spin-1/2 Kitaev model. At $H\neq 0$, the Wilson loop $W_l=\langle \hat{W}_l \rangle $ measured at the ground state is shown in Fig.~\ref{Fig:PhaseDiagram} (d),  where ${W}_y$ takes the exact quantized value at $H=0$. After the time reversal symmetry breaking by the magnetic filed, ${W}_y$ takes nearly quantized values for a finite range of magnetic fields at $H<H_*^\text{AFM}$ ($H<H_*^\text{FM}$) for AFM (FM) Kitaev coupling. This indicates the $\mathbb{Z}_2$ gauge structure in the pure S=1 Kitaev model remains a good description of perturbed Kitaev model away from the zero-field limit.
Here we should note $W_y>0$ for $L_y=3$ and $W_y<0$ for $L_y=4$, while its absolute value is nearly quantized. This indicates distinct topological sectors can be accessed with changing $L_y$. Beyond $H_*^\text{FM}$ or $H_{*}^\text{AFM}$, ${W}_y$ tends to be negligibly small, particularly for wider systems, implying the absence of $\mathbb{Z}_2$ gauge structure. The sudden drop in ${W}_y$ characterizes the phase transition at $H_*^\text{AFM}$ and $H_*^\text{FM}$.

\emph{The KSL Phase.}---Below we show that the phase in Fig.~\ref{Fig:PhaseDiagram}(a) at $0<H<H_*^\text{AFM}$ in AFM model or $0<H<H_*^\text{FM}$ in FM model is a gapped KSL phase with topological nature. We first examine the spin-spin correlations $\mathbb{S}_{ij}\equiv\langle \mathbf{S_i}\cdot \mathbf{S_j}\rangle-\langle \mathbf{S_i} \rangle\cdot\langle\mathbf{S_j}\rangle$ and the Von Neumann entanglement entropy $S_\text{vN}$ to show the gapped nature, and then we probe its topological properties from the Wilson loop, the entanglement spectra and the edge modes.

In the Kitaev limit (i.e., H=0),  we confirm that the spin-spin correlations vanish beyond the nearest-neighbor links~\cite{Baskaran} like $S=1/2$ pure Kitaev model~\cite{Kitaev2006}. Beyond the Kitaev limit, as shown in Figs.~\ref{Fig:GappedSL} (a) for both AFM and FM Kitaev couplings, the spin-spin correlations decay exponentially along the zigzag chains of the cylinders [see the inset of  Figs.~\ref{Fig:GappedSL} (a)], indicating a nonmagnetic ground state with finite spin gap. The gapped nature can be further confirmed by the Von Neumann entanglement entropy ${S_\text{vN}}=-\text{Tr}\left( {{\rho _\text{A}}\ln {\rho _\text{A}}} \right)$, where ${\rho_\text{A}}$ is the reduced density matrix of part $A$ for the bipartition of the system into $A$ and $B$, with ${\rho_\text{A}}$ obtained by tracing out the degrees of freedom of the $B$ part. Here, we consider the cut parallel to $\mathbf{e_y}$ and measure $S_\text{vN}$ for each cut with subsystem length  $L_A$.  As shown in Fig.~\ref{Fig:GappedSL} (b), we find that $S_\text{vN}$ are independent on the positions of each cut and display flat behavior for cylinders with different widths, suggesting zero central charge or the existence of finite energy gap in the bulk.

In addition, we also examine the entanglement spectra for the cut  parallel to $\mathbf{e_y}$ at $L_A=L_x/2$, as shown in Fig.~\ref{Fig:GappedSL} (c), with a finite entanglement gap $\Delta_\xi$ separating the ground-state manifold from the higher spectra. On cylinders of width $L_y=4$, the entanglement spectra displays fourfold (quasi-)degeneracy [see Fig.~\ref{Fig:GappedSL} (c)], indicating the existence of boundary zero modes and its topological nature. The structure  of entanglement spectra is similar to the topological KSL in spin S=1/2 Kitaev model~\cite{ZZ2018}. The topological nature is also exhibited by the Wilson loop operator $\hat{W}_y$, which takes nearly quantized mean value in the KSL phase [see Fig.~\ref{Fig:PhaseDiagram} (c)]. Moreover, the distribution of ${W_y}$  along the cylinders is uniform with the nearly quantized value [see inset of Fig.~\ref{Fig:GappedSL} (d)]. All of these facts imply the $\mathbb{Z}_2$ gauge structure remains a good description of the gapped KSL phase. To explicitly show the presence of edge modes, we compute the local on-site  magnetization distributions $\langle\mathbf{S}\cdot\mathbf{e_{\text{H}}}\rangle$ [$\mathbf{e_{\text{H}}}=(1,1,1)$] along the zigzag chain of the cylinders, as shown in Fig.~\ref{Fig:GappedSL} (d), where the spins near the edge are much easier to be polarized than the bulk, indeed implying the gapless edge modes.

\begin{figure}[!t]
\begin{center}
\includegraphics[width=0.5\textwidth]{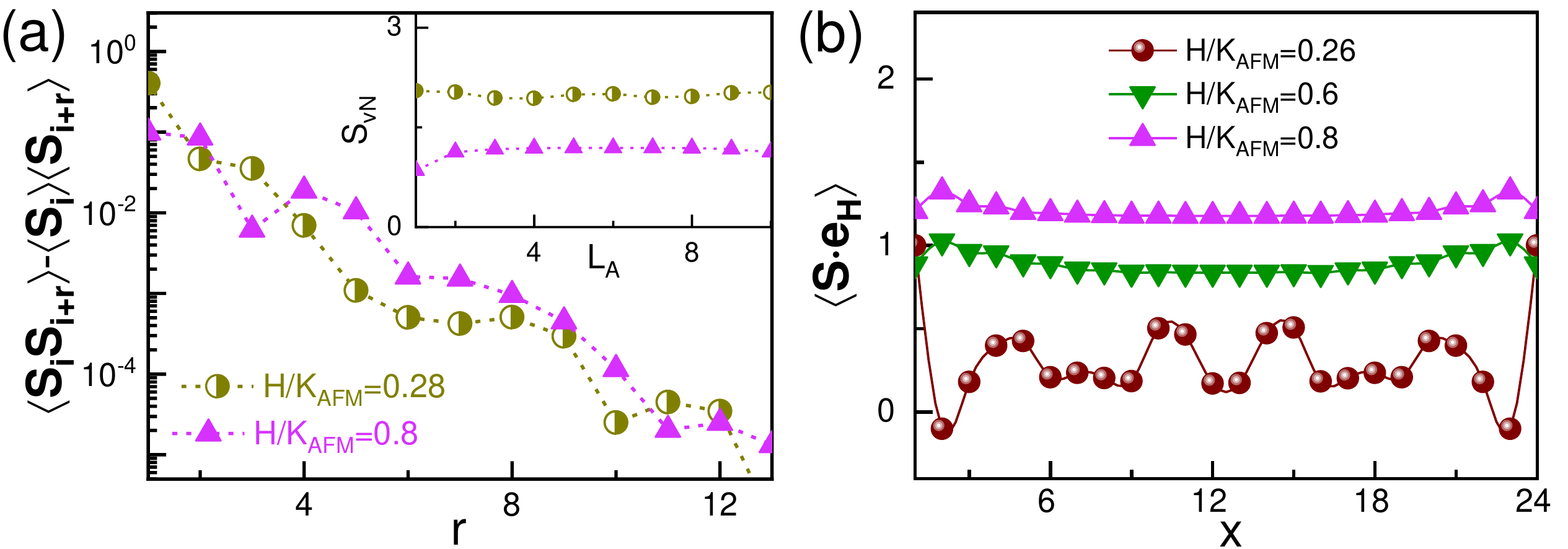}
\end{center}
\par
\renewcommand{\figurename}{Fig.}
\caption{(Color online) Intermediate phase (IP) and paramagnet phase in the AFM Kitaev model. Panel (a) shows the decay of spin correlations along the zigzag chain in semi-logarithmic scale for $L_y=3, L_x=12$ cylinders, where we compared the intermediate phase with the paramagnets. The inset of (a) exhibits the flat behavior of the entanglement entropy $S_\text{VN}$ as a function of the cut position $L_A$. Panel (b) shows the spatial distribution of $\langle\mathbf{S}\cdot\mathbf{e_{\text{H}}}\rangle$ [$\mathbf{e_{\text{H}}}=(1,1,1)$] along the zigzag chain. }
\label{Fig:IP}
\end{figure}

From the above analysis, the nearly quantized Wilson loop operator mean-value, the four-fold degeneracy in the entanglement spectra  and the existence of edge modes all manifest the topological nature of the gapped KSL phase.
Different from the FM model, the susceptibility $\chi$ in Fig.~\ref{Fig:PhaseDiagram} (b) suggests the AFM model further hosts an intermediate phase sandwiched between the KSL and partially polarized paramagnet phase at $H_*^\text{AFM}<H<H_{**}^\text{AFM}$. The intermediate phase is distinct from KSL phase by the vanishingly small Wilson loop,  the disappearance of the degeneracy in entanglement spectra together with the edge mode. As shown in Fig.~\ref{Fig:IP} (a) and its inset, $\mathbb{S}_{ij}$ decays exponentially while $S_\text{vN}$ is flat.
However,  the local magnetization $\langle\mathbf{S}\cdot\mathbf{e_{\text{H}}}\rangle$ shows strong oscillation in real space, which is similar to the intermediate spin liquid phase in the S=1/2 case. Due to the nonuniform pattern of magnetization, much larger system sizes are demanded to resolve its nature, which is beyond the scope of this work.

\emph{Instability of the KSL.}---We have found that the KSL in the AFM Kitaev model is much more robust against increasing magnetic field than the one in the FM model (i.e., $H_*^\text{AFM}\gg H_*^\text{FM}$), though at zero field these two models are equivalent~\cite{Chaloupka2010}. For finite magnetic fields, the indications of such significant difference can be gained from the low-field magnitude of susceptibility $\chi$ [see Fig.~\ref{Fig:PhaseDiagram} (c)].

\begin{figure}[t]
\begin{center}
\includegraphics[width=0.4\textwidth]{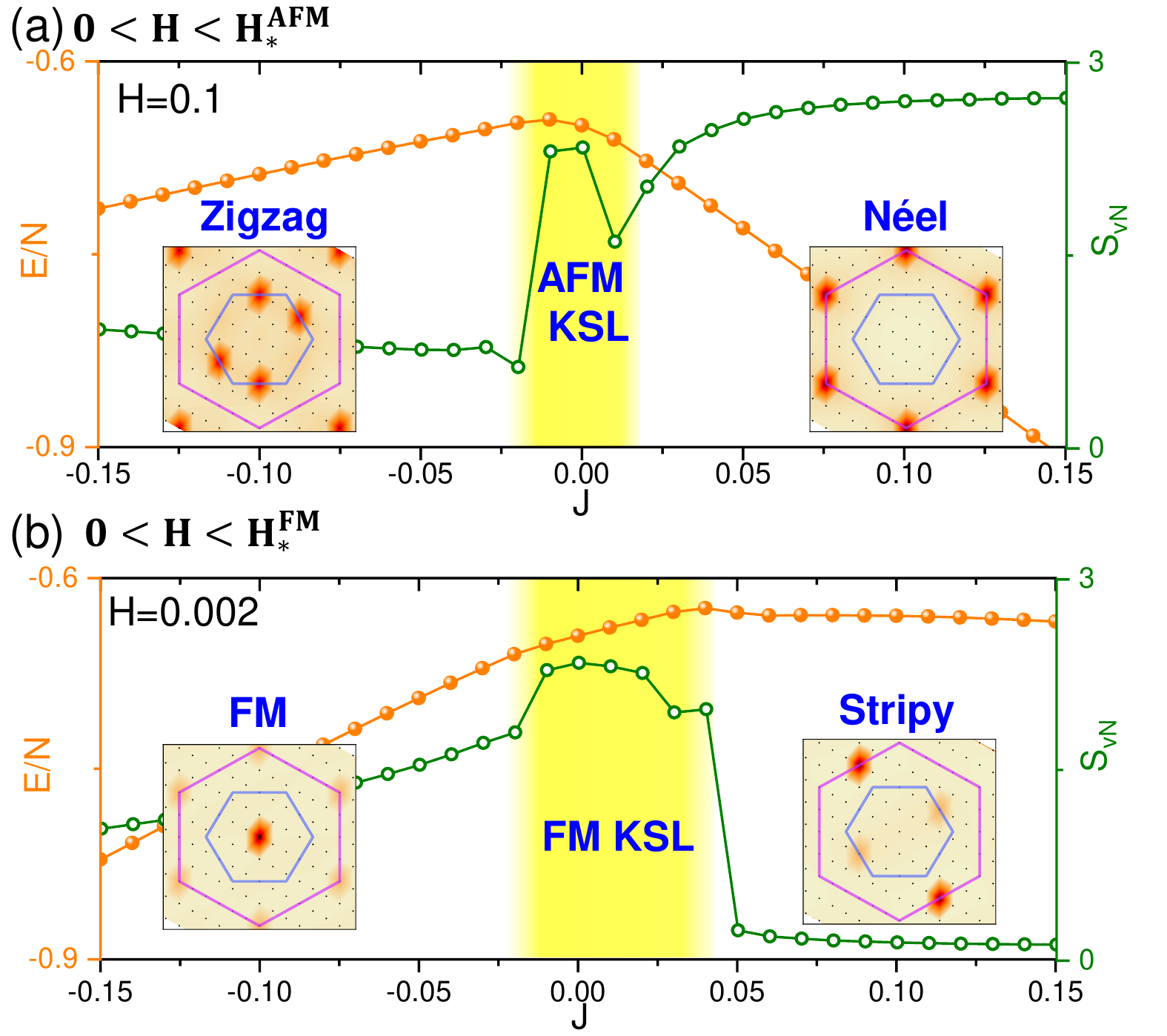}
\end{center}
\par
\renewcommand{\figurename}{Fig.}
\caption{(Color online) Instability of the KSL. Panels (a-b) show the instability of the AFM KSL (a) and FM KSL (b) against AFM/FM Heisenberg perturbations. The colored shadows correspond to the KSL depicted in Fig.~\ref{Fig:PhaseDiagram}. The ground state energy per site (orange color) and the entanglement entropy for the cuts bisecting the torus (olive color) show that the AFM/FM KSLs survive in a narrow range of Heisenberg coupling strength and eventually give way to magnetic orders. Here we consider torus geometry with 4x3 unit cells. {The nature of these magnetic orders are identified by the static spin structure factor (the insets) at $|J/K|=0.1$. The black dots represent the accessible momenta in the Brillouin zone for finite sized cylinders  with $L_y=4, L_x=6$.}}
\label{Fig:KHmodel}
\end{figure}

Furthermore, in the S=1 Kitaev materials, the superexchange processes would also give rise to the Heisenberg interactions between nearest-neighbor sites~\cite{Stavropoulos2019}. Thus we study the instability of the KSL against the Heisenberg interactions in the vicinity of the pure Kitaev model.
Here we consider both AFM and FM Heisenberg coupling $J$. The phase transitions are characterized by the singular behavior of entanglement entropy and the kinks in the ground-state energy density. As illustrated in the Figs.~\ref{Fig:KHmodel} (a-b), the magnetic field induced gapped KSLs can survive in a finite range of the Heisenberg interactions before they finally give way to the magnetic ordered phases. These magnetic ordered phases are identified by the peaks in the static spin structure factor,
, as shown in the insets of Figs.~\ref{Fig:KHmodel}. Here the magnetic ordered phases at $H>0$  are the same as the ones identified in the Kitaev-Heisenberg model with the same ratio $J/K$ but $H=0$~\cite{Stavropoulos2019,DMRG}. In other words, if the system is tuned proximate to the Kitaev phase at a finite $J$, the spin liquids we have discovered can be still induced with applying a magnetic field along [111] direction.

 \emph{Discussions and Summary.}---
We investigate the magnetic field induced phases in the S=1 Kitaev honeycomb model, which is distinct from the studies on the pure S=1 Kitaev model in Refs.~\cite{Baskaran,Minakawa2019,ED,TPQ,DMRG,Lee2019} because the magnetic field breaks the time reversal symmetry and gives rise to new spin liquids. The FM and AFM models are no longer equivalent beyond zero-field limit either. The gapped topological KSL identified in this work is indeed different from the gapless KSL proposed in the pure Kitaev model~\cite{Baskaran,Minakawa2019,ED,TPQ,DMRG}.

We identify a field-induced gapped KSL phase with topological nature for both the AFM and FM Kitaev couplings, including $\mathbb{Z}_2$ gauge structure, the edge modes and degenerate entanglement spectra on cylinders. Interestingly, such KSL phase is much more stable against increasing magnetic field for the AFM Kitaev coupling than the FM one, which resembles to the S=1/2 case~\cite{ZZ2018}. However, there are differences between S=1 and S=1/2 cases such as the degeneracy in the entanglement spectrum. Moreover, the interaction type of S=1/2 Kitaev materials is proposed to be FM~\cite{Trebst2017,Hermanns2017} while the S=1 Kitaev materials harbor AFM Kitaev interactions~\cite{Stavropoulos2019}.

Theoretical understanding of S=1 Kitaev model in a magnetic field  is still a challenge partially due to the lack of exact solution even in the zero-field limit, and thus it will be an excellent task to pursue in the future. Our findings may also stimulate the investigations on the effect of more complicated interactions or anisotropies, as well as the theories of the intermediate phase.
Recently, the  S=1 and high-spin Kitaev materials~\cite{Stavropoulos2019,Xu2020} are proposed to host AFM Kitaev couplings, then it is promising to realize stable topological spin liquids based on our findings. The signatures such as the quantized thermal Hall conductivity and fractionalized excitations in inelastic neutron scattering can be directly probed.

\begin{acknowledgments}
Note added:  At the final stage of finishing this work, we noticed that Ref.\cite{Lee2019} constructed a tensor network ground-state wavefunction for the pure S=1 Kitaev model.  The magnetic field effect was also examined in Ref.\cite{Lee2019} via the magnetizations, which are consistent with our results in Fig.~\ref{Fig:PhaseDiagram} (b-c). In our work, we identify the magnetic field induced AFM/FM KSL from different aspects of systematical characterization based on unbiased DMRG method. We should also note that the largest field for the AFM model in Ref.\cite{Lee2019} is up to $H\approx0.3/\sqrt{3}<H_*^\text{AFM}$, which lies within the AFM KSL phase identified in our work. After submitting our paper, we also noticed that Ref.\cite{Khait2020,Hickey2020,Koga2020} also report the investigations on the S=1 Kitaev model in a magnetic field.

\emph{Acknowledgments}--- Z.Z. would like to thank the discussions with Ashvin Vishwanath. We acknowledge the computational resources at
CSUN and KITS for performing the numerical simulations in this work.
This material is based upon work supported by the U.S. Department of Energy,
Office of Basic Energy Sciences under the grant No. DE-FG02-06ER46305 (D.N.S).
Z.Z. appreciatively acknowledges funding via Ashvin Vishwanath at Harvard University.
Z.-Y. W. is supported by Natural Science Foundation of China (Grant No. 11534007) and MOST of China (Grant No. 2017YFA0302902).
\end{acknowledgments}

\end{document}